# Light localization in nonuniformly randomized lattices

Yaroslav V. Kartashov,[1] Vladimir V. Konotop,[2] Victor A. Vysloukh,[1] and Lluis Torner[1]

[1]ICFO-Institut de Ciencies Fotoniques, and Universitat Politecnica de Catalunya, Mediterranean Technology Park, 08860 Castelldefels (Barcelona), Spain
[2]Centro de Fisica Teorica e Computacional and Departamento de Fisica, Faculdade de Ciencias, Universidade de Lisboa, Avenida Professor Gama Pinto 2, Lisboa 1649-003, Portugal

We address Anderson localization of light in disordered optical lattices where the disorder strength varies across the transverse direction. Such variation changes the preferred domains where formation of localized eigenmodes is most probable, hence drastically impacting light localization properties. Thus, step-like disorder results in formation of modes with different decay rates at both sides of the interface, while a smoothly varying disorder yields appearance of modes that are extended within weakly disordered domains and rapidly fade away in strongly disordered domains.

Wave and particle scattering in disordered materials are topics of continuously renewed interest. Such scattering may exhibit transition from a ballistic regime to complete transport suppression, termed Anderson localization [1]. Optics affords a unique laboratory for the exploration of such phenomena [2-4]. Anderson localization has been observed in optically-induced [5] and fabricated [6-8] uniform lattices. While in unbounded disordered samples all eigenmodes would be localized, the presence of boundaries drastically affects light propagation [7,9-12]. Under suitable conditions, disorder can even enhance wave transport in finite systems instead of arresting it [13]. In this context, upon analysis of Anderson light localization the disorder strength is usually considered invariable across the sample [6-14]. In contrast, in this Letter we study Anderson localization of light in non-uniformly randomized systems and show that if the *statistical properties* of disorder depend on the transverse coordinate, a specific *statistical inhomogeneity* emerges that drastically affects the light evolution.

We address light propagation in one-dimensional disordered waveguide arrays that can be described by the nonlinear Schrödinger equation for the field amplitude $q$:

$$i\frac{\partial q}{\partial \xi} = -\frac{1}{2}\frac{\partial^2 q}{\partial \eta^2} - R(\eta)q - q|q|^2. \quad (1)$$

Here $\eta$ and $\xi$ are the normalized transverse and longitudinal coordinates, respectively. The array shape is described by the function $R(\eta) = \sum_m p_m G(\eta - \eta_m)$, where individual waveguides $G(\eta) = \exp(-\eta^6/a^6)$ feature the width $a$ and refractive index $p_m$. The coordinates of the waveguide centers $\eta_m = md + r_m$ are randomized, with $d$ being the regular spacing and $r_m$ being a random shift of the waveguide position uniformly distributed within $[-S_m, S_m]$. We study *spatially inhomogeneous disorder*, i.e. $S_m$ depends on the waveguide position. In contrast, the array depth $p_m$ is always regular, and it can take two different values at $m \geq 0$ and $m < 0$ if an interface is considered. Such arrays can be fabricated with laser-writing techniques [7]. We solved Eq. (1) with a split-step Fourier method, for $d = 1.6$, which corresponds to a waveguide spacing of 16 $\mu$m, and $a = 0.3$, which corresponds to a width of 3 $\mu$m. We report results averaged over $n_{tot} = 10^3$ realizations, but accurate data are obtained with a few dozen realizations, as may be the case in practice. The value $p_m \sim 11$ corresponds to refractive index contrast $\delta n \sim 1.3 \times 10^{-3}$ at $\lambda = 800$ nm.

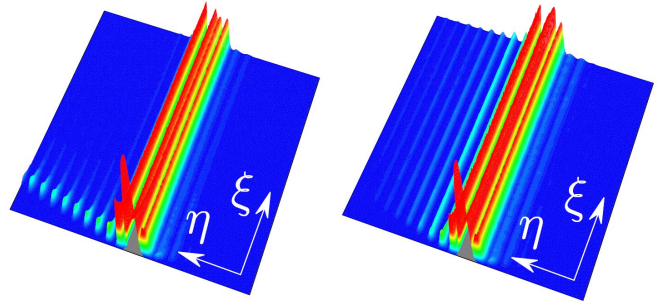

Figure 1. Dynamics of propagation up to $\xi = 600$ averaged over $10^3$ disorder realizations at $S_l = 0.0$ (left) and $S_l = 0.2$ (right). In both cases $S_r = 0.6$ and $p_l = p_r = 11$.

First we address step-like disorder by introducing the concept of statistical interface between two arrays defined by different disorder levels, $S_m = S_l$ at $m < 0$ and $S_m = S_r$ at $m \geq 0$, and/or by different refractive indexes $p_m = p_l$ at $m < 0$ and $p_m = p_r$ at $m \geq 0$. We integrate Eq. (1) for different disorder realizations and for the input $q|_{\xi=0} = Aw(\eta)$ corresponding to the linear guided mode $w(\eta)$ with the amplitude $A$ of an isolated waveguide, with either $m = -1$ or $m = 0$ (i.e. the mode was launched at the statistical interface). Figure 1 shows the ensemble-averaged light intensity distributions, i.e. $I(\eta, \xi) = n_{tot}^{-1}\sum_{n=1}^{n_{tot}} |q_n(\eta, \xi)|^2$, where $n$ is a realization number, corresponding to different propagation scenarios at $p_l = p_r$ in the linear case, i.e. at $A \to 0$. The disorder in the right array was set to $S_r = 0.6$, while disorder in the left array $S_l$ increased gradually. When $S_l = 0$ (left panel) one observes the formation of statistical interface waves between disordered and regular arrays [see Fig. 2(a) for averaged intensity distribution at $\xi = 600$]. They decay exponentially in the right array and exhibit a long tail after an interval of fast exponential decay in the left array. To exclude reflections from the boundaries at $\eta \to \pm\infty$ in the simulations the arrays dimensions were set to far exceed the width of regular discrete diffraction pattern at $\xi = 600$.

The different decay types are related to different phenomena at both sides of the interface: while in the right array decay is determined by the disorder level, in the left array one can distinguish modes having the propagation constant in the gap of the periodic spectrum, which contribute to the formation of localized surface states, and modes whose propagation constants belong to allowed bands and which move away from the surface. The statistical weights of each type of modes are determined by the particular realization of the right array. When $S_l \neq 0$, the diffraction tail disappears and localization in the left array also becomes exponential but with smaller exponent for $S_l < S_r$ [Fig. 2(b)]. The exponential decay of the averaged output intensity distribution in both arrays becomes equal at $S_l = S_r$ [Fig. 2(c)]. To quantify the impact of disorder on the degree of localization, we studied the decay rates $\alpha_l, \alpha_r$ in the left and right arrays using different linear fits of $\ln[I(\eta)]$ at $\eta < 0$ and $\eta > 0$. The dependences of $\alpha_l, \alpha_r$ on the disorder strength $S_l$ in the left array for $S_r = 0.6$ are shown in Fig. 3(a).

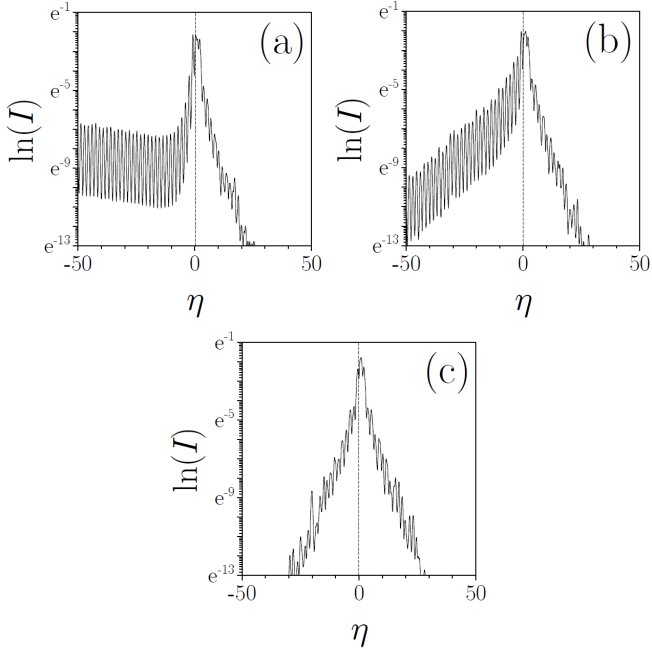

Figure 2. Average output intensity distributions at (a) $S_l = 0.0$, (b) $S_l = 0.2$, and (c) $S_l = 0.6$ obtained at $\xi = 600$ for $10^3$ disorder realizations. In all cases $S_r = 0.6$ and $p_l = p_r = 11$.

Surprisingly, while $\alpha_l$ increases almost linearly with $S_l$, the decay rate in the right array $\alpha_r$ is affected by disorder in the left array, especially for $S_l \geq 0.4$. Thus, the statistical interface affects the degree of localization at both sides of the interface and modifies the decay rates that otherwise would be obtained in the equivalent uniformly-disordered array. The difference in localization introduced by the statistical interface can be understood by comparing the localization degree for $S_l = 0$, $S_r \neq 0$ and $S_l = S_r \neq 0$. In the former case only the modes associated to the gap of the regular array contribute to the formation of localized interface wave, i.e. the number of realizations where interface modes appear is reduced due to the presence of the regular array.

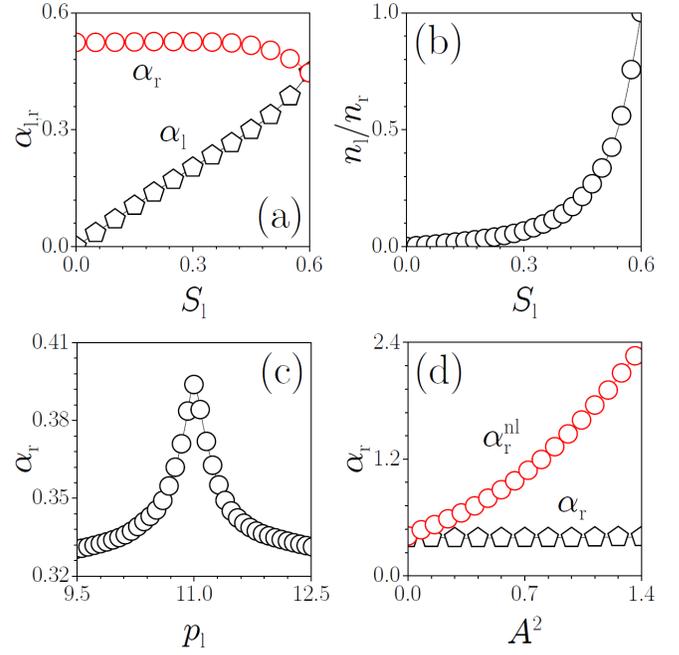

Figure 3. (a) Decay rates $\alpha_l, \alpha_r$ versus $S_l$ at $p_l = p_r = 11$ and $S_r = 0.6$. (b) $n_l/n_r$ versus $S_l$ at $p_l = p_r = 11$ and $S_r = 0.6$. (c) Decay rate $\alpha_r$ versus $p_l$ at $S_l = 0.0$, $S_r = 0.4$, $p_r = 11$. (d) Decay rates $\alpha_r^{nl}, \alpha_r$ versus $A^2$ at $S_l = 0.0$, $S_r = 0.4$, $p_l = p_r = 11$.

We calculated the stationary eigenmodes of array with step-like disorder, using the equation $bQ = -(1/2)\partial^2 Q/\partial\eta^2 - R(\eta)Q$ with zero boundary conditions that can be obtained from Eq. (1) upon substitution $q(\eta,\xi) = Q(\eta)\exp(-ib\xi)$ in linear case. We used 31 waveguides since the size of the array is limited by the calculation time required to solve the linear eigenvalue problem. We found the integral center position $\eta_c = U^{-1}\int_{-\infty}^{\infty} \eta Q^2 d\eta$, where $U = \int_{-\infty}^{\infty} Q^2 d\eta$, of most localized fundamental mode with largest $b$ (whose shape is usually not affected by the array boundaries) for each realization and counted the number of events $n_r$ when $\eta_c > 0$ (mode localized in the right array) and $n_l$ when $\eta_c < 0$ (mode localized in the left array) for $n_{tot} = 10^3$ realizations $(n_l + n_r = n_{tot})$.

The ratio $n_l/n_r$ rapidly grows with $S_l$ at fixed $S_r$ and approaches unity for $S_l = S_r$ [Fig. 3(b)] when the fundamental mode can form in any point of the array with equal probability. Such modes usually form on two waveguides with the smallest separation. The appearance of such pairs of closely spaced waveguides is more probable in domains with large disorder. This is supported by the histograms of random position of the integral center $\eta_c$ of the fundamental eigenmode [Figs. 4(a),(b)]. Such histograms do not decay with $m$, i.e. in a given array the eigenmodes are almost uniformly distributed between the statistical interface at $m = 0$ and the array boundary at $m = \pm 15$. This result is similar to the absence of a sharp transition between surface and bulk modes in finite regular arrays [15].

A difference in the refractive index of the two arrays also affects the light localization. Figure 3(c) shows the decay rate $\alpha_r$ of the averaged intensity distribution in the right array versus the refractive index $p_l$ in the left array, for $p_r = 11$ and step-like disorder. While introduction of a mismatch between refractive indices $p_l, p_r$ may even reduce the rate of discrete diffraction in each array, it also makes the interface repulsive (as in truncated arrays [7]) which reduces the localization degree at a fixed disorder level. Then, the decay rate $\alpha_r$ takes its maximal value at $p_l = p_r$ in the absence of additional internal ($p_l > p_r$) or external ($p_l < p_r$) reflection at the interface of regular and disordered arrays.

The impact of nonlinearity on light localization in arrays with step-like disorder is illustrated in Fig. 3(d). A focusing cubic nonlinearity modifies the averaged intensity distributions mostly around their peaks, where nonlinearity is strong, and weakly affects the tails of such distributions where $q \to 0$. Thus, a fitting of the averaged intensity distributions can be performed using double-exponent function, where one exponent describes the decay rate $\alpha_r^{nl}$ in the vicinity of launching point, while other exponent captures the decay rate $\alpha_r$ of low-amplitude tails. The decay rate $\alpha_r^{nl}$ affected by the nonlinearity rapidly grows with increase of input peak amplitude $A$, while the decay rate of low-amplitude tails is independent of $A$.

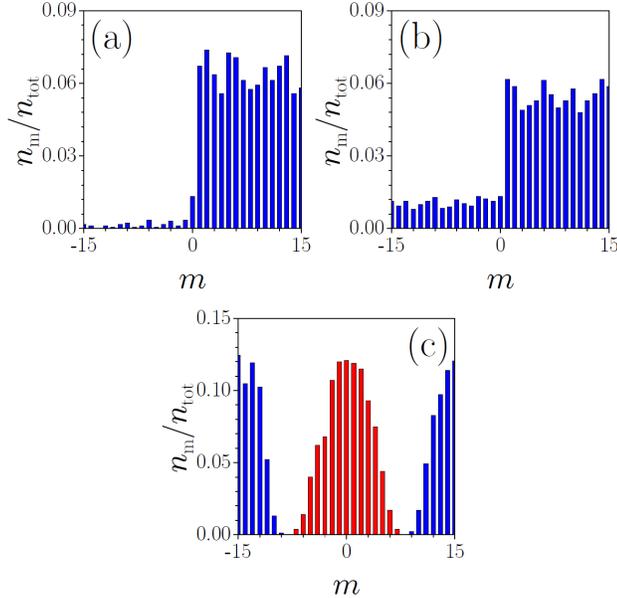

Figure 4. Histograms showing the ratio of the number of events $n_m$ when the maximum of the lowest-order eigenmode is localized on $m$-th waveguide to $n_{\text{tot}}$ for step-like disorder with (a) $S_l = 0.3$, $S_r = 0.6$ and (b) $S_l = 0.4$, $S_r = 0.6$ and (c) disorder growing to the periphery (blue) and to the center (red) of array with $\max(S_{l,r}) = 0.5$. In all cases $p_l = p_r = 11$.

We also considered spatial variations of the disorder strength described by the function $S_m = S_{\max} \exp(-m^2/k^2)$, where $k = 8$ stands for the width of the randomized domain. The single-peak histogram depicted in Fig. 4(c) indicates the high probability of the fundamental mode excitation in such domain. In the opposite case of randomization on the array periphery $S_m = S_{\max}[1 - \exp(-m^2/k^2)]$ the highly localized modes appear more frequently near the array boundaries [blue-colored histogram in Fig. 4(c)]. The corresponding averaged output intensity distributions are shown in Fig. 5. In arrays randomized in the center, light is localized in the vicinity of the launching point, but outside the disordered region a radiative background appears [Fig. 5(a)]. In contrast, in arrays disordered at the periphery, the output intensity distribution is extended within weakly disordered domain, but it rapidly decays in strongly disordered arrays [Fig. 5(b)]. If disorder varies linearly in the transverse plane one observes localized averaged intensity distributions, where the decay rate changes continuously with $\eta$ [Fig. 5(c)].

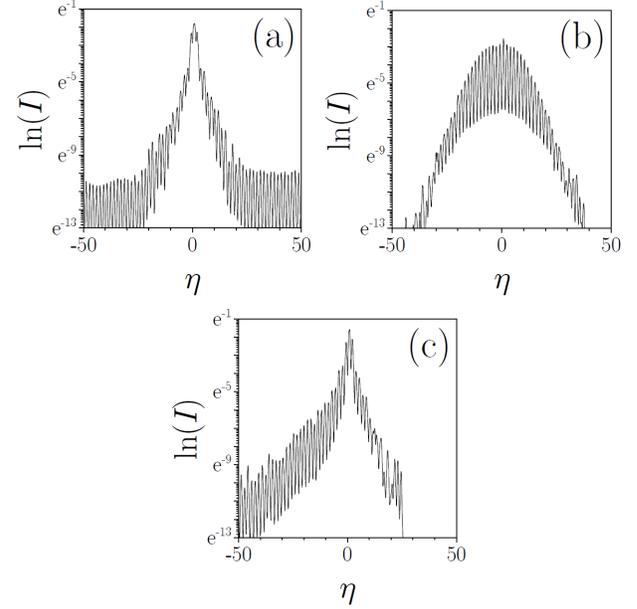

Figure 5. Average output intensity distributions obtained at $\xi = 600$ for $n_{\text{tot}} = 10^3$ in the case when (a) disorder strength decreases from $0.5$ in the center of array to $0$ at its periphery, (b) when disorder strength grows from 0 at the periphery of the array to 0.5 in its center, and (c) when disorder strength increases linearly from 0.2 to 0.6 within 10 waveguides. In all cases $p_l = p_r = 11$.

### References

1. P. W. Anderson, Phys. Rev. **109**, 1492 (1958).
2. H. de Raedt, A. Lagendijk, and P. de Vries, Phys. Rev. Lett. **62**, 47 (1989).
3. A. A. Chabanov, M. Stoytchev, and A. Z. Genack, Nature **404**, 850 (2000).
4. T. Pertsch, U. Peschel, J. Kobelke, K. Schuster, H. Bartelt, S. Nolte, A. Tünnermann, and F. Lederer, Phys. Rev. Lett. **93**, 053901 (2004).
5. T. Schwartz, G. Bartal, S. Fishman and M. Segev, Nature **466**, 52 (2007).
6. Y. Lahini, A. Avidan, F. Pozzi, M. Sorel, R. Morandotti, D. N. Christodoulides, and Y. Silberberg, Phys. Rev. Lett. **100**, 013906 (2008).
7. A. Szameit, Y. V. Kartashov, P. Zeil, F. Dreisow, M. Heinrich, R. Keil, S. Nolte, A. Tünnermann, V. A. Vysloukh, and L. Torner, Opt. Let. **35**, 1172 (2010).
8. L. Martin, G. Di Giuseppe, A. Perez-Leija, R. Keil, F. Dreisow, M. Heinrich, S. Nolte, A. Szameit, A. F. Abouraddy, D. N. Christodoulides, and B. E. A. Saleh, Opt. Express **19**, 13636 (2011).
9. D. M. Jovic, Y. S. Kivshar, C. Denz, and M. R. Belic, Phys. Rev. A **83**, 033813 (2011).
10. V. Folli and C. Conti, Opt. Lett. **36**, 2830 (2011).
11. F. Lederer, G. I. Stegeman, D. N. Christodoulides, G. Assanto, M. Segev, and Y. Silberberg, Phys. Rep. **463**, 1 (2008).
12. Y. V. Kartashov, V. A. Vysloukh, and L. Torner, Prog. Opt. **52**, 63 (2009).
13. Y. V. Kartashov, V. A. Vysloukh, and L. Torner, Opt. Let. **36**, 466 (2011).
14. S. Ghosh, G. P. Agrawal, B. P. Pal, and R. K. Varshney, Opt. Comm. **284**, 201 (2011).
15. Y. V. Bludov, V. V. Konotop, Phys. Rev. E **76**, 046604 (2007).